\documentclass[12pt]{article}
\usepackage{amsmath} 
\input{epsf} 
\newcommand\as{\alpha_{\mathrm{S}}} 
\newcommand\f[2]{\frac{#1}{#2}} 
\def\beq{\begin{equation}} 
\def\eeq{\end{equation}} 
\def\beeq{\begin{eqnarray}} 
\def\eeeq{\end{eqnarray}} 
 
\def\to{\rightarrow}
\def\ito{\leftarrow} 
\def\nn{\nonumber}

\def\ms{${\overline {\rm MS}}$}

\begin{document} 
\begin{titlepage}
\begin{flushright}
ZU-TH 16/12
\end{flushright}
\renewcommand{\thefootnote}{\fnsymbol{footnote}}
\vspace*{2cm}

\begin{center}
{\Large \bf Vector boson production at hadron colliders:\\[0.3cm]
hard-collinear coefficients at the NNLO}
\end{center}

\par \vspace{2mm}
\begin{center}
{\bf Stefano Catani${}^{(a)},$ Leandro Cieri${}^{(a)},$
Daniel de Florian${}^{(b)}$, Giancarlo Ferrera${}^{(c)},$\\
}
and
{\bf Massimiliano Grazzini${}^{(d)}$\footnote{On leave of absence from INFN, Sezione di Firenze, Sesto Fiorentino, Florence, Italy.}}\\

\vspace{5mm}

$^{(a)}$ INFN, Sezione di Firenze and Dipartimento di Fisica e Astronomia,\\ 
Universit\`a di Firenze,
I-50019 Sesto Fiorentino, Florence, Italy\\

${}^{(b)}$Departamento de F\'\i sica, FCEYN, Universidad de Buenos Aires,\\
(1428) Pabell\'on 1 Ciudad Universitaria, Capital Federal, Argentina\\

${}^{(c)}$Dipartimento di Fisica, Universit\`a di Milano and\\ INFN, Sezione di Milano,
I-20133 Milan, Italy\\

$^{(d)}$ Institut f\"ur Theoretische Physik, Universit\"at Z\"urich, CH-8057 Z\"urich, Switzerland

\vspace{5mm}

\end{center}

\par \vspace{2mm}
\begin{center} {\large \bf Abstract} \end{center}
\begin{quote}
\pretolerance 10000

We consider QCD radiative corrections to vector-boson production in hadron collisions.
We present the next-to-next-to-leading order (NNLO) result
of the hard-collinear coefficient function 
for the all-order resummation of logarithmically-enhanced
contributions at small transverse momenta.
The coefficient function controls NNLO contributions 
in resummed calculations at full next-to-next-to-leading logarithmic accuracy.
The same coefficient function is 
used
in applications of the subtraction method to perform 
fully-exclusive perturbative calculations up to NNLO.

\end{quote}

\vspace*{\fill}
\begin{flushleft}
August 2012

\end{flushleft}
\end{titlepage}

\setcounter{footnote}{1}
\renewcommand{\thefootnote}{\fnsymbol{footnote}}

The transverse-momentum $(q_T)$
distribution of systems with high invariant mass $M$
(Drell-Yan lepton pairs, vector boson(s), Higgs boson(s) and so forth)
produced in hadronic collisions is important for physics studies within
and beyond the Standard Model (SM).

The computation of these distributions in perturbative QCD is complicated by the presence of
large logarithmic contributions of the form $\ln (M^2/q_T^2)$ that need to be
resummed to all perturbative orders in the QCD coupling $\as$. 
The method to perform the resummation is known 
\cite{Dokshitzer:hw, Collins:1981uk, Collins:1984kg, Catani:2000vq}, 
including recent developments on the discovered and resummed effects 
\cite{Nadolsky:2007ba, Catani:2010pd}
due to helicity and azimuthal correlations in gluon fusion subprocesses.
The structure of the resummed calculation is organized in a process-independent
form that is controlled by a set of perturbative functions with computable 
`resummation coefficients'.
All the resummation coefficients that are process independent are known 
since some time
\cite{Kodaira:1981nh,Davies:1984hs,Catani:vd,deFlorian:2001zd} 
up to the second order in $\as$, and 
the third-order coefficient $A^{(3)}$ has been obtained in 
Ref.~\cite{Becher:2010tm}. The complete computations of the second-order
resummation coefficients have been carried out in 
Refs.~\cite{Catani:2007vq} and \cite{Catani:2009sm} for
two benchmark processes,
namely, the production of the SM Higgs boson through  gluon fusion and vector
boson production through the Drell--Yan (DY) mechanism of quark--antiquark
annihilation.
The explicit analytic expressions for the ${\cal O}(\as^2)$
hard-collinear resummation coefficients 
in the case of SM Higgs boson production in the large-$m_{top}$ limit 
have been presented in Ref.~\cite{Catani:2011kr}. This paper
parallels Ref.~\cite{Catani:2011kr}:
we concentrate on single vector boson production,
and we present the corresponding analytic expressions of the 
second-order hard-collinear coefficient functions ${\cal H}^{(2)}$.

QCD predictions for vector boson production at hadron colliders are important
for present and forthcoming studies at the Tevatron and the LHC.
Resummed calculations of the $q_T$ spectrum of vector bosons and of related
observables are presented in 
Refs.~\cite{Balazs:1997xd}--\cite{Banfi:2012yh}.
Calculations for vector boson production at the fully-exclusive level with
respect to the accompanying QCD radiation have been carried out in 
Refs.~\cite{Melnikov:2006di, Catani:2009sm, Gavin:2010az}
up to the next-to-next-to-leading order (NNLO) in perturbative QCD.

In this paper we compute the hard-collinear coefficient 
function ${\cal H}^{(2)}$ and, thus, the complete analytical expression of the
NNLO cross section for vector boson production in the small-$q_T$ region.
These results have a twofold relevance, in the context of both resummed and
fixed-order calculations.

The knowledge of ${\cal H}^{(2)}$ can be implemented in resummed calculations
at full next-to-next-to-leading logarithmic (NNLL) order to achieve uniform NNLO
accuracy in the small-$q_T$ region. In the case of vector boson production,
this implementation has been carried out in Ref.~\cite{Bozzi:2010xn}
by using the impact-parameter space resummation formalism developed in
Refs.~\cite{Bozzi:2005wk, Bozzi:2007pn}. This formalism enforces 
a unitarity constraint and thus it guarantees that (upon inclusion of 
${\cal H}^{(2)}$) the resummed $q_T$ spectrum returns the complete NNLO total
cross section after integration over $q_T$. 

The subtraction method of  Ref.~\cite{Catani:2007vq} exploits the 
knowledge of transverse-momentum resummation coefficients at 
${\cal O}(\as^2)$ to perform NNLO calculations at the fully-exclusive level.
The Higgs boson coefficient functions presented in Ref.~\cite{Catani:2011kr}
were used in the numerical computations of 
Refs.~\cite{Catani:2007vq, Grazzini:2008tf}.
The coefficient functions presented in this paper 
are precisely those that are needed
for the actual implementation of this subtraction method 
in DY-type processes: they are used in Refs.~\cite{Catani:2009sm} 
and \cite{Ferrera:2011bk} for the NNLO numerical computations of 
vector boson production and of associated production 
of a Higgs boson and a $W$ boson. The diphoton NNLO calculation of 
Ref.~\cite{Catani:2011qz} also uses part of the results of the present paper to
treat the quark-antiquark annihilation subprocess $q{\bar q} \to \gamma \gamma$.

This paper is organized as follows. We first introduce our notation and
illustrate the NNLO calculation of the vector boson cross section at small
values of $q_T$. Then we recall the transverse-momentum resummation formalism.
Finally, we present our NNLO results in analytic form and the relation with the
$q_T$ resummation coefficients at ${\cal O}(\as^2)$.

We briefly introduce the theoretical framework and our notation.
We consider the production of a vector boson $V$ 
($V=W^\pm,Z$ and/or $\gamma^*$)
in hadron--hadron collisions.
We use the narrow width approximation and we treat the vector boson as an
on-shell particle with mass $M$.
The QCD expression of the vector boson transverse-momentum
cross section\footnote{If $V=\gamma^*$ or if the vector boson $V$ is not an on-shell particle, 
the transverse-momentum cross section $d\sigma/d q_T^2$ has to be replaced by the
doubly-differential 
distribution $M^2d\sigma/d M^2 d q_T^2\,$, where $M$ is the invariant mass of $V$.} is
\begin{equation}
\label{dcross}
\f{d\sigma}{d q_T^2}(q_T,M,s)=\sum_{a,b}
\int_0^1 dz_1 \,\int_0^1 dz_2 \,f_{a/h_1}(z_1,M^2)
\,f_{b/h_2}(z_2,M^2) \;
\f{d{\hat \sigma}_{ab}}{d q_T^2}(q_T, M,{\hat s}=z_1z_2s; \as(M^2)) 
\;,
\end{equation}
where $f_{a/h_i}(x,\mu_F^2)$ ($a=q_f,{\bar q_f},g$) are the parton densities of 
the colliding hadrons ($h_1$ and $h_2$) at the factorization scale $\mu_F$,
and $d{\hat \sigma}_{ab}/d q_T^2$ are the
partonic cross sections. The centre--of--mass energy of the two colliding
hadrons is denoted by $s$, and ${\hat s}$ is the partonic centre--of--mass
energy. We use parton
densities as defined in the \ms\
factorization scheme, and $\as(\mu_R^2)$ is the QCD running coupling 
at the renormalization scale $\mu_R$ in the \ms\
renormalization scheme. In Eq.~(\ref{dcross}) and throughout the paper,
the arbitrary factorization and renormalization scales, $\mu_F$ and $\mu_R$,
are set to be equal to the vector boson mass $M$.

The partonic cross sections $d{\hat \sigma}_{ab}/d q_T^2$ are computable in
QCD perturbation theory as power series expansions in $\as(M^2)$.
We are interested in the perturbative contributions that are large in the 
small-$q_T$ region $(q_T \ll M)$ and, eventually, singular in the limit 
$q_T \to 0$.
To explicitly 
recall
the perturbative structure of these enhanced terms at
small $q_T$, we follow Ref.~\cite{Catani:2011kr}
and we introduce the cumulative partonic cross section\footnote{In our notation, the subscripts $c$ and ${\bar c}$ denote a quark 
and an antiquark (or viceversa) that do not necessarily have 
the same flavour. The flavour structure depends on the produced 
vector boson $V$ and it is (implicitly) specified by the specific 
form of the Born level cross section $\sigma^{(0)}_{c {\bar c},V}$. }
\begin{equation}
\label{inte}
\int_0^{Q_0^2}dq_T^2 
\;\f{d{\hat \sigma}_{ab}}{dq_T^2}(q_T,M,{\hat s}=M^2/z;\as(M^2))
\equiv \sum_{c=q_f, \,{\bar q}_{f'}} z \; \sigma_{c {\bar c},V}^{(0)} 
\;{\hat R}_{c {\bar c} \ito ab}^V(z,M/Q_0;\as(M^2)) \;\;,
\end{equation}
where the overall normalization of the function ${\hat R}^V$ is defined with
respect to $\sigma_{q_f {\bar q}_{f'}, V}^{(0)}$,
which is the Born level cross section for the 
quark--antiquark annihilation
subprocess $q_f{\bar q}_{f'}\to V$
(the quark flavours $f$ and $f'$ are equal if $V=Z, \gamma^*$).
The partonic function ${\hat R}^V$ has the following perturbative expansion
\begin{equation}
\label{eqnR}
{\hat R}_{c {\bar c} \ito ab}^V(z,M/Q_0;\as)=\delta_{ca} \,\delta_{\bar c b} \,\delta(1-z)+
\sum_{n=1}^\infty
\left(\f{\as}{\pi}\right)^n\, {\hat R}^{V (n)}_{c {\bar c} \ito ab}(z,M/Q_0) \;\; .
\end{equation}
The next-to-leading order (NLO) and NNLO contributions to the cumulative cross
section in Eq.~(\ref{inte}) are determined by the 
functions ${\hat R}^{V (1)}$ and
${\hat R}^{V (2)}$, respectively.
The small-$q_T$ region of the 
cross section
$d{\hat \sigma}_{ab}/d q_T^2$ is probed by performing the limit $Q_0\ll M$
in Eq.~(\ref{inte}). In this limit, the NLO and NNLO functions 
${\hat R}^{V (1)}$ and ${\hat R}^{V (2)}$ have the following
behaviour:
\begin{equation}
\label{eqr1s}
{\hat R}^{V (1)}_{c {\bar c} \ito ab}(z,M/Q_0)=l_0^2 \;{\hat R}_{c {\bar c} \ito ab}^{(1; 2)}(z)
+l_0 \;{\hat R}_{c {\bar c} \ito ab}^{(1; 1)}(z)
+{\hat R}_{c {\bar c} \ito ab}^{(1; 0)}(z)+{\cal O}(Q_0^2/M^2) \;\;,
\end{equation}
\begin{align}
\label{eqr2s}
{\hat R}^{V (2)}_{c {\bar c} \ito ab}(z,M/Q_0)
&= l_0^4\; {\hat R}_{c {\bar c} \ito ab}^{(2;4)}(z)
+l_0^3\; {\hat R}_{c {\bar c} \ito ab}^{(2;3)}(z)+l_0^2\; 
{\hat R}_{c {\bar c} \ito ab}^{(2;2)}(z) \nn \\ 
&+ l_0 \;{\hat R}_{c {\bar c} \ito ab}^{(2;1)}(z)
+ {\hat R}_{c {\bar c} \ito ab}^{(2;0)}(z)
+{\cal O}(Q_0^2/M^2) \;,
\end{align}
where $l_0=\ln (M^2/Q_0^2)$. In Eqs.~(\ref{eqr1s}) and (\ref{eqr2s}), the powers
of the large logarithm $l_0$ are produced by the singular (though, integrable)
behaviour of $d{\hat \sigma}_{ab}/d q_T^2$ at small values of $q_T$.
The coefficients ${\hat R}^{(1;m)}$ (with $m \leq 2$) and ${\hat R}^{(2;m)}$
(with $m \leq 4$) of the large logarithms are independent of $Q_0$; these
coefficients depend on the partonic centre--of--mass energy $\hat s$ and,
more precisely, they are functions of the energy fraction $z=M^2/\hat s$.
As is well known (see also Eq.~(\ref{qtycrossgg})), the logarithmic coefficients 
${\hat R}^{(n;m)}$ do not depend on the specific vector boson that is produced by 
$q\bar q$ annihilation and, therefore, we have removed the explicit superscript $V$
(i.e., ${\hat R}^{V (n;m)}= {\hat R}^{(n;m)}$).

In this paper we present the result of  the computation of the cumulative cross section in
Eq.~(\ref{inte}) up to NNLO. The partonic calculation is performed in analytic
form by neglecting terms of ${\cal O}(Q_0^2/M^2)$ in the limit $Q_0 \ll M$.
Therefore, we determine the coefficient functions ${\hat R}^{(n;m)}(z)$
in Eqs.~(\ref{eqr1s}) and (\ref{eqr2s}). 

To perform our calculation, we follow the same method as used in 
Ref.~\cite{Catani:2011kr} to evaluate the transverse-momentum cross section for Higgs
boson production. The $q_T$ integration in Eq.~(\ref{inte}) is thus rewritten in the
following form:
\begin{align}
\label{master}
\int_0^{Q_0^2}dq_T^2 
\;\f{d{\hat \sigma}_{ab}}{dq_T^2}(q_T,M,{\hat s};\as) 
&
\equiv \int_0^{+\infty} dq_T^2 
\;\f{d{\hat \sigma}_{ab}}{dq_T^2}(q_T,M,{\hat s};\as)
-\int_{Q_0^2}^{+\infty} dq_T^2 
\;\f{d{\hat \sigma}_{ab}}{dq_T^2}(q_T,M,{\hat s};\as)
\nn \\
&=
{\hat \sigma}_{ab}^{({\rm tot})}(M,{\hat s};\as)
-\int_{Q_0^2}^\infty dq_T^2 \int^{+\infty}_{-\infty} d{\hat y}
\;\f{d{\hat \sigma}_{ab}}{d{\hat y} \,dq_T^2}({\hat y},q_T,M,{\hat s};\as) 
\;\;,
\end{align}
where ${\hat \sigma}_{ab}^{({\rm tot})}$ is the vector boson total (i.e. integrated
over $q_T$) cross section and $d{\hat \sigma}_{ab}/d{\hat y} \,dq_T^2$ is the
corresponding doubly-differential cross section with respect to the
transverse momentum and rapidity ($\hat y$ is the rapidity of $V$ in the
centre--of--mass frame of the two colliding partons $a$ and $b$) of the vector boson.
The total cross section ${\hat \sigma}_{ab}^{({\rm tot})}(M,{\hat s};\as)$ is known 
\cite{Hamberg:1990np} in analytic form up to NNLO (i.e., up to ${\cal O}(\as^2 
\sigma_{V}^{(0)})$). In the region of large or, more precisely, non-vanishing 
values of $q_T$, the differential distribution 
$d{\hat \sigma}_{ab}/d{\hat y} \,dq_T^2$ is also known 
\cite{Ellis:1981hk,Gonsalves:1989ar,Arnold:1988dp} in analytic form up to 
${\cal O}(\as^2 \sigma_{V}^{(0)})$. Using these known results and exploiting
Eq.~(\ref{master}), we can compute the cumulative partonic cross section up to the
NNLO. Note that $q_T > Q_0$ 
in the last term on the right-hand side of Eq.~(\ref{master}). Therefore the
corresponding integration of the expression
$d{\hat \sigma}_{ab}/d{\hat y} \,dq_T^2$ \cite{Ellis:1981hk,Gonsalves:1989ar,Arnold:1988dp}
over ${\hat y}$ and $q_T^2$ is finite as long as $Q_0\neq 0$: using the explicit 
expression of $d{\hat \sigma}_{ab}/d{\hat y} \,dq_T^2$ 
from\footnote{
We list some typos that we have found and corrected in some formulae of
Ref.~\cite{Gonsalves:1989ar}. In Eq.~(2.12), $B_2^{qG}$ has to be replaced by
$B_2^{qG}+C_2^{qG}$, and $C_2^{qG}$ has to be replaced by $C_3^{qG}$.
In Eq.~(A.4), two signs have to be changed:
$B_1^{qG}$ has to be replaced by $- B_1^{qG}$,  
and $A^{qG}$ has to be replaced by $- A^{qG}$.
In the first line of Eq.~(A.10), the term $C_F\,(f_u-f_s-f_t)$ has 
to be replaced by $C_A\,(f_u-f_s-f_t)$.} 
Ref.~\cite{Gonsalves:1989ar},
we carry out the integration in analytic from in the
limit $Q_0 \ll M$ (i.e., we neglect terms
of ${\cal O}(Q_0^2/M^2)$ on the right-hand side of Eq.~(\ref{master})).
The result of our calculation\footnote{Some technical details related to the limit
$Q_0 \ll M$ are illustrated in Ref.~\cite{Catani:2011kr}.} confirms the logarithmic
structure in Eqs.~(\ref{eqr1s}) and (\ref{eqr2s}), and it allows us to determine the
NLO and NNLO coefficients 
${\hat R}^{(1;m)}$ (with $m \leq 2$) and ${\hat R}^{(2;m)}$
(with $m \leq 4$) of the cumulative cross section in Eq.~(\ref{inte}).

The results of the coefficient functions ${\hat R}^{(n;m)}(z)$ are conveniently
expressed in terms of transverse-momentum resummation coefficients.
Therefore, before presenting the results, we
recall how these functions are
related to the perturbative coefficients of the transverse-momentum resummation
formula for vector boson production \cite{Collins:1984kg}.
This relation
also shows that from the knowledge of Eq.~(\ref{eqr2s})
we can fully determine the NNLO {\em rapidity} distribution
of the vector boson in the small-$q_T$ region.

To present the transverse-momentum resummation formula, we first decompose
the partonic cross section $d{\hat \sigma}_{ab}/d q_T^2$ in Eq.~(\ref{dcross})
in the form 
$d{\hat \sigma}_{ab} = d{\hat \sigma}_{ab}^{({\rm sing})} 
+ d{\hat \sigma}_{ab}^{({\rm reg})}$.
The singular component, $d{\hat \sigma}_{ab}^{({\rm sing})}$,
contains all the contributions that are enhanced at small $q_T$.
These contributions are proportional to
$\delta(q_T^2)$ or to large logarithms
of the type 
$1/q_T^2\,\ln^m (M^2/q_T^2)$.
The remaining component, $d{\hat \sigma}_{ab}^{({\rm reg})}$,
of the partonic cross section is regular 
order-by-order in $\as$ as $q_T \to 0$: the
integration of $d{\hat \sigma}_{ab}^{({\rm reg})}/dq_T^2$ over the range 
$0 \leq q_T \leq Q_0$ leads to a result that, at each
fixed order in $\as$, {\em vanishes} in the limit $Q_0 \to 0$.
Therefore, $d{\hat \sigma}_{ab}^{({\rm reg})}$ only contributes to the terms
of ${\cal O}(Q_0^2/M^2)$ on the right-hand side of 
Eqs.~(\ref{eqr1s}) and (\ref{eqr2s}).

Inserting the decomposition $d{\hat \sigma}_{ab} = d{\hat \sigma}_{ab}^{({\rm sing})} 
+ d{\hat \sigma}_{ab}^{({\rm reg})}$ in Eq.~(\ref{dcross}), we obtain a 
corresponding decomposition,
$d\sigma = d\sigma^{({\rm sing})} 
+ d\sigma^{({\rm reg})}$,
of the hadronic cross section.
The transverse-momentum resummation formula for the singular component
of the $q_T$ cross section at fixed value of the rapidity $y$
(the rapidity is defined in the centre--of--mass frame of the two 
colliding hadrons) of the vector boson reads
\cite{Collins:1984kg, Catani:2000vq}
\beeq
\label{qtycrossgg}
&&\!\!\!\!\!\!\f{{d\sigma}^{({\rm sing})}}{dy \,d q_T^2}(y,q_T,M,s) =
\f{M^2}{s} \sum_{c=q_f, \,{\bar q}_{f'}}
\sigma_{c {\bar c},V}^{(0)} 
\int_0^{+\infty} db \;\f{b}{2}  \;J_0(b q_T) \;
  S_q(M,b)\nn \\
&& \;\;\;\; \times \;
\sum_{a_1,a_2} \,
\int_{x_1}^1 \f{dz_1}{z_1} \,\int_{x_2}^1 \f{dz_2}{z_2} 
\; \left[ H^F C_1 C_2 \right]_{c \bar c;\,a_1a_2}
\;f_{a_1/h_1}(x_1/z_1,b_0^2/b^2)
\;f_{a_2/h_2}(x_2/z_2,b_0^2/b^2) \;
\;, 
\eeeq
where the kinematical variables $x_i$ $(i=1,2)$ are 
$x_1= e^{+y} M/{\sqrt s}$ and $x_2= e^{-y} M/{\sqrt s}$. The integration
variable $b$ is the impact parameter, $J_0(b q_T)$ is the $0$th-order 
Bessel function, and $b_0=2e^{-\gamma_E}$
($\gamma_E=0.5772\dots$ is the Euler number) is a numerical coefficient.
The symbol $\left[ H^F C_1 C_2 \right]_{c \bar c;\,a_1a_2}$ shortly denotes
the following function of the longitudinal-momentum fractions
$z_1$ and $z_2$:
\beq
\label{whath}
\left[ H^{DY} C_1 C_2 \right]_{c \bar c;\,a_1a_2}
= H_{q}^{DY}(\as(M^2)) 
\;\; C_{c \,a_1}(z_1;\as(b_0^2/b^2)) 
\;\; C_{{\bar c} \,a_2}(z_2;\as(b_0^2/b^2)) 
\;\;, 
\eeq
where $H_{q}^{DY}(\as)$ and $C_{c \,a}(z;\as)$ ($c=q_f, {\bar q}_f$)
 are
perturbative functions of $\as$ (see Eqs.~(\ref{cpert})--(\ref{hpert})).

The quark form factor $S_q(M,b)$ in Eq.~(\ref{qtycrossgg}) is a
process-independent quantity \cite{Collins:1984kg, Kodaira:1981nh, Catani:2000vq}. 
Its functional
dependence on $M$ and $b$ is controlled by two perturbative functions, which are
usually denoted as $A_q(\as)$ and $B_q(\as)$
(see, e.g., Ref.~\cite{Catani:2010pd} that uses the same notation as in
Eq.~(\ref{qtycrossgg})). 
Their corresponding $n$-th
order perturbative coefficients are $A_q^{(n)}$ and $B_q^{(n)}$. The
coefficients $A_q^{(1)}$, $B_q^{(1)}$, $A_q^{(2)}$ \cite{Kodaira:1981nh} and
$B_q^{(2)}$ \cite{Davies:1984hs} are known: their knowledge fully determines
the perturbative expression of $S_q(M,b)$ up to ${\cal O}(\as^2)$.

\setcounter{footnote}{1}

The perturbative function $H_{q}^{DY}(\as)$ in Eq.~(\ref{whath})
is process dependent, since it is directly related to the production mechanism 
of the vector boson through quark--antiquark annihilation. However, 
$H_{q}^{DY}$
is independent of the specific type of vector boson $V$ ($V=W^\pm, Z,
\gamma^*$), and we have introduced the generic superscript $DY$.

The partonic functions $C_{q_f \,a}$ and $C_{{\bar q}_f \,a}$ 
in Eq.~(\ref{whath})
are instead process independent, as a consequence of the universality 
features of QCD collinear radiation.
Owing to their process independence, these partonic functions
fulfil the following relations:
\beeq
\label{fdep1}
C_{q_f \,q_{f'}}(z;\as) = C_{{\bar q}_f \,{\bar q}_{f'}}(z;\as) 
\equiv C_{q \, q}(z;\as) \;\delta_{f f'} + C_{q \, q'}(z;\as)\;
\left(1 - \delta_{f f'} \right)\;\;, \\
\label{fdep2}
C_{q_f \,{\bar q}_{f'}}(z;\as) = C_{{\bar q}_f \,{q}_{f'}}(z;\as) 
\equiv C_{q \, \bar q}(z;\as) \;\delta_{f f'} + C_{q \, \bar{q}'}(z;\as)\;
\left(1 - \delta_{f f'} \right)\;\;, 
\\
\label{fdep3}
C_{q_f \,g}(z;\as) = C_{{\bar q}_f \,g}(z;\as) \equiv C_{q \, g}(z;\as)\;\;,
\qquad \qquad \qquad \qquad\;\;\;\;\;\;
\eeeq
which are a consequence of charge conjugation invariance and 
flavour symmetry of QCD. The dependence of the matrix  $C_{c\,a}$ on the parton
labels is thus fully specified by the five independent quark functions  
$C_{q\,q}$, $C_{q\,q'}$, $C_{q\,{\bar q}}$, $C_{q\,{\bar q}'}$ and $C_{q\,g}$
on the right-hand side of Eqs.~(\ref{fdep1})--(\ref{fdep3}). 

We recall that the function 
$H_{q}^{DY}(\as)$, the quark functions  $C_{q \,a}(\as)$
and the perturbative function
$B_q(\as)$ of the quark form factor are not {\em separately} computable in an
unambiguous way. Indeed, these three functions are related by a
renormalization-group symmetry \cite{Catani:2000vq} that 
follows from the 
$b$-space factorization structure of Eq.~(\ref{qtycrossgg}). 
The unambiguous definition of these three functions thus requires the
specification\footnote{The reader who is not interested in issues related to the
specification of a resummation scheme can simply assume that
$H_{q}^{DY}(\as) \equiv 1$ throughout this paper. The choice
$H_{q}^{DY}(\as) = 1$ is customarily used in most of the literature
on $q_T$ resummation for vector boson production.}
of a {\em resummation scheme} \cite{Catani:2000vq}.
Note, however, that considering the perturbative 
expansion\footnote{The resummation-scheme dependence also cancels
by consistently expanding Eq.~(\ref{qtycrossgg}) in terms of classes of
resummed (leading, next-to-leading and so forth) logarithmic contributions
\cite{Bozzi:2005wk}.}
of Eq.~(\ref{qtycrossgg}) (i.e., the perturbative expansion of the singular
component of the $q_T$ cross section), the resummation-scheme dependence exactly
cancels order-by-order in $\as$. 

The perturbative expansion of the quark functions $C_{q \,a}(\as)$ and of the
vector boson function $H_{q}^{DY}(\as)$
is defined as follows:
\begin{equation}
\label{cpert}
C_{q\,a}(z; \as)=\delta_{q\,a} \;\delta(1-z)+
\sum_{n=1}^{\infty}\left(\frac{\as}{\pi}\right)^n C_{q\,a}^{(n)}(z) \;\; ,
\quad (a=g,q,\bar q, q',\bar {q}') \;,
\end{equation}
\begin{equation}
\label{hpert}
H_q^{DY}(\as) = 1+\sum_{n=1}^\infty \left(\frac{\as}{\pi}\right)^n H_q^{DY(n)}\, .
\end{equation}
The first-order coefficient function $C_{q\,g}^{(1)}(z)$ is independent of
the resummation scheme; its expression is \cite{Davies:1984hs}
\beq
\label{coneqg}
C_{q\,g}^{(1)}(z) = \f{1}{2}\;z\;(1-z)  \;\;. 
\eeq
The first-order coefficients 
$C_{q\,q'}^{(1)}(z)$, $C_{q\,{\bar q}}^{(1)}(z)$ and $C_{q\,{\bar q}'}^{(1)}(z)$
vanish,
\beq
\label{cone0}
C_{q\,q'}^{(1)}(z)=C_{q\,{\bar q}}^{(1)}(z)= C_{q\,{\bar q}'}^{(1)}(z)=0\;\;,
\eeq
while the coefficients
$C_{q\,q}^{(1)}(z)$ and $H_q^{DY(1)}$ fulfill the following relation
\cite{Davies:1984hs, deFlorian:2001zd}:
\beq
\label{coneqq}
C_{q\,q}^{(1)}(z) + \f{1}{2} \,H_q^{DY(1)} \,\delta(1-z) 
=  \f{C_F}{2}\left(\bigg(\f{\pi^2}2-4\bigg)\,\delta(1-z)+1-z\right)\;\;.
\eeq
The separate determination of $C_{q\,q}^{(1)}(z)$ and $H_q^{DY(1)}$ requires the
specification of a resummation scheme. For instance, considering the 
resummation scheme in which the coefficient $H_q^{DY(1)}$
vanishes, the right-hand side of Eq.~(\ref{coneqq}) gives the value of 
$C_{q\,q}^{(1)}(z)$,
and the corresponding value 
of the quark form factor coefficient $B_q^{(2)}$ is explicitly computed
in Ref.~\cite{Davies:1984hs}.
The computation of the second-order coefficients 
$C_{q\,q}^{(2)}$, $C_{q\,q'}^{(2)}$, $C_{q\,{\bar q}}^{(2)}$, $C_{q\,{\bar q}'}^{(2)}$, $C_{q\,g}^{(2)}$
and 
$H_q^{DY(2)}$ is the 
aim of the calculation
described in this paper.

To the purpose of presenting the NNLO results for the cumulative cross section
in Eq.~(\ref{inte}),
we also define the following hard-collinear coefficient
function:
\begin{equation}
\label{HCCGG}
{\cal H}^{DY}_{q \bar q\ito ab}(z;\as) \equiv H_q^{DY}(\as) \!\int_0^1\!dz_1 \int_0^1\!dz_2
\,\delta(z - z_1z_2)
\, C_{q \,a}(z_1;\as)\, C_{{\bar q} \,b}(z_2;\as)\, ,
\end{equation}
which is directly related to the coefficient function in Eq.~(\ref{whath}).
The function ${\cal H}^{DY}$ depends only on the energy fraction $z$,
and it arises after integration  
of the resummation formula (\ref{qtycrossgg})
over the rapidity of the vector boson.
Note that ${\cal H}^{DY}$ is 
independent of the resummation scheme \cite{Catani:2000vq}.
The perturbative expansion of the function
${\cal H}^{DY}$ directly follows from Eqs.~(\ref{cpert})--(\ref{hpert}). We have
\begin{equation}
\label{chpert}
{\cal H}^{DY}_{q \bar q \ito ab}(z;\as) =\delta_{q\,a} \,\delta_{\bar q\,b}\;\delta(1-z)+
\sum_{n=1}^{\infty}\left(\frac{\as}{\pi}\right)^n 
{\cal H}^{DY(n)}_{q \bar q\ito ab}(z) \;\; ,
\end{equation}
where the first-order and second-order contributions are
\begin{equation}
\label{H1}
{\cal H}^{DY(1)}_{q \bar q\ito ab}(z)=\delta_{q\,a} \,\delta_{\bar q\,b} \,\delta(1-z) 
\,H^{DY(1)}_q+
\delta_{q\,a} \,C^{(1)}_{\bar q\,b}(z)+\delta_{\bar q\,b} \,C^{(1)}_{q\,a}(z) \;\; ,
\end{equation}
\begin{align}
\label{H2}
{\cal H}^{DY(2)}_{q \bar q\ito ab}(z)&=\delta_{q\,a} \,\delta_{\bar q\,b} \,\delta(1-z)
\,H^{DY(2)}_q
+\delta_{q\,a} \,C^{(2)}_{\bar q\,b}(z)+\delta_{\bar q\,b} \,C^{(2)}_{q\,a}(z)
\nn\\
&
+H^{DY(1)}_q\left(\delta_{q\,a} \,C^{(1)}_{\bar q\,b}(z)
+\delta_{\bar q\,b} \,C^{(1)}_{q\,a}(z)\right)
+\left(C^{(1)}_{q\,a}\otimes C^{(1)}_{\bar q\,b}\right)(z)\;\; .
\end{align}
In Eq.~(\ref{H2}) and in the following, the symbol $\otimes$ denotes the 
convolution integral
(i.e., we define 
$(g \otimes h) (z) \equiv \int_0^1 dz_1 \int_0^1 dz_2
\,\delta(z - z_1z_2) \;g(z_1) \;h(z_2)$).

\setcounter{footnote}{2}

In the limit $Q_0 \ll M$,
the perturbative expansion of the cumulative partonic cross section in Eq.~(\ref{inte})
can directly be related to the resummation coefficients of Eq.~(\ref{qtycrossgg}).
We refer the reader to Ref.~\cite{Catani:2011kr} for a concise illustration of
this relation and to Ref.~\cite{Bozzi:2005wk} for more technical details.
The NLO and NNLO functions 
${\hat R}^{V (1)}$ and ${\hat R}^{V (2)}$ in Eqs.~(\ref{eqr1s}) and (\ref{eqr2s})
have the following
expressions:
\begin{equation}
\label{eqr1}
{\hat R}^{V (1)}_{q \bar q \ito ab}(z,M/Q_0)=l_0^2 \;\Sigma_{q\bar q \ito ab}^{DY(1;2)}(z)
+l_0\,\Sigma_{q\bar q \ito ab}^{DY(1;1)}(z)+{\cal H}_{q\bar q\ito ab}^{DY(1)}(z)
+{\cal O}(Q_0^2/M^2) \;\;,
\end{equation}
\begin{align}
\label{eqr2}
\!{\hat R}^{V (2)}_{q\bar q \ito ab}(z,M/Q_0)&=l_0^4 \,\Sigma_{q\bar q\ito ab}^{DY(2;4)}(z)
+l_0^3\, \Sigma_{q\bar q\ito ab}^{DY(2;3)}(z)+l_0^2\, \Sigma_{q \bar q\ito ab}^{DY(2;2)}(z)
+l_0 \Big(\Sigma_{q\bar q\ito ab}^{DY(2;1)}(z)-16 \zeta_3 \Sigma_{q\bar q\ito ab}^{DY(2;4)}(z)\Big)
\!\nn\\
&
+\left({\cal H}_{q\bar q\ito ab}^{DY(2)}(z)
-4\zeta_3\, \Sigma_{q\bar q\ito ab}^{DY(2;3)}(z)\right)
+{\cal O}(Q_0^2/M^2)\, ,
\end{align}
where we have used the same notation
as in Ref.~\cite{Bozzi:2005wk}. The explicit expressions of the coefficient functions 
$\Sigma_{q\bar q\ito ab}^{DY(n;m)}(z)$ in terms of the resummation coefficients are given
in Eqs.~(63),(64),(66)--(69) of Ref.~\cite{Bozzi:2005wk}
(we have to set $\mu_R=\mu_F=Q=M$, where $\mu_R,\, \mu_F$ and $Q$
are the auxiliary scales of Ref.~\cite{Bozzi:2005wk})
and are not reported here.
The coefficients
${\cal H}_{q\bar q\ito ab}^{DY(1)}$ and ${\cal H}_{q\bar q\ito ab}^{DY(2)}$ are exactly 
those in Eqs.~(\ref{H1}) and (\ref{H2})
(they are also given in Eqs.~(65) and (70) of Ref.~\cite{Bozzi:2005wk})
The first-order terms 
$\Sigma_{q\bar q\ito ab}^{DY(1;2)}$ and $\Sigma_{q\bar q\ito ab}^{DY(1;1)}$ depend
on the quark form factor $S_q(M,b)$. The second-order terms
$\Sigma_{q\bar q\ito ab}^{DY(2;m)}$ depend on 
${\cal H}_{q\bar q\ito ab}^{DY(1)}$ and on the quark form factor $S_q(M,b)$
up to ${\cal O}(\as^2)$. The numerical coefficient 
$\zeta_3 \simeq 1.202\dots$ ($\zeta_k$ is the Riemann $\zeta$-function) 
on the right-hand side of Eq.~(\ref{eqr2}) originates from the Bessel
transformations 
(see, e.g., Eqs.~(B.18) and (B.30)
in Appendix~B of Ref.~\cite{Bozzi:2005wk}).


We now document our results of the NNLO computation of the cumulative partonic
cross section.
Using Eqs.~(\ref{eqr1}) and (\ref{eqr2}), the results for ${\hat R}^{V (1)}$
and ${\hat R}^{V (2)}$
allow us to extract 
$\Sigma^{DY(n;m)}$ and ${\cal H}^{DY(n)}$ up to ${\cal O}(\as^2)$.
The explicit result of the NLO function ${\hat R}^{V (1)}(z)$ confirms the
expressions of $\Sigma_{q\bar q\ito ab}^{DY(1;2)}(z)$,  
$\Sigma_{q\bar q\ito ab}^{DY(1;1)}(z)$ and ${\cal H}^{DY(1)}_{q\bar q\ito ab}(z)$, as
predicted by the $q_T$ resummation coefficients at ${\cal O}(\as)$.
At NNLO, the present knowledge \cite{Kodaira:1981nh, Davies:1984hs} of the 
$q_T$ resummation coefficients at ${\cal O}(\as^2)$ predicts the expressions of
the terms $\Sigma_{q\bar q\ito ab}^{DY(2;m)}(z)$, with $m=1,2,3,4$. Our result for the
NNLO function ${\hat R}^{V (2)}(z)$ confirms this prediction, and it allows us
to extract the explicit expression of the second-order coefficient function 
${\cal H}^{DY(2)}_{q\bar q\ito ab}(z)$.

\newpage
We obtain
\begin{align}
\label{h2qqb}
{\cal H}^{DY(2)}_{q\bar q\ito q \bar q}(z)&=
C_A C_F 
\bigg\{
\left(\frac{7 \zeta_3}{2}-\frac{101}{27}\right)\left(\frac{1}{1-z}\right)_+ 
+\left(\frac{59 \zeta_3}{18}-\frac{1535}{192}+\frac{215 \pi ^2}{216}-\frac{\pi ^4}{240}\right) \delta(1-z)
\nn\\
&~~~~~~
+\frac{1+z^2}{1-z}
 \bigg(-\frac{\text{Li}_3(1-z)}{2}+\text{Li}_3(z)-\frac{\text{Li}_2(z) \log (z)}{2}
-\frac{1}{2} \text{Li}_2(z) \log (1-z)-\frac{1}{24} \log ^3(z)
\nn\\
&~~~~~~
-\frac{1}{2} \log ^2(1-z) \log (z)+\frac{1}{12} \pi ^2 \log (1-z)-\frac{\pi ^2}{8}\bigg)
+\frac{1}{1-z} \bigg(
-\frac{1}{4} \left(11-3 z^2\right) \zeta_3
\nn\\
&~~~~~~
-\frac{1}{48} \left(-z^2+12 z+11\right) \log ^2(z)-\frac{1}{36} \left(83 z^2-36 z+29\right) \log (z)+\frac{\pi ^2 z}{4}\bigg)
\nn\\
&~~~~~~
+(1-z) \bigg(\frac{\text{Li}_2(z)}{2}+\frac{1}{2} \log (1-z) \log (z)\bigg)
+\frac{z+100}{27}+\frac{1}{4} z \log (1-z)\bigg\}
\nn\\
&~~~~~~
+C_F n_F \bigg\{\frac{14}{27}\left(\frac{1}{1-z}\right)_+ 
+\frac{1}{864} \left(192 \zeta_3+1143-152 \pi ^2\right) \delta(1-z)
\nn\\
&~~~~~~
+\frac{\left(1+z^2\right)}{72 (1-z)} \log (z) (3 \log (z)+10)+\frac{1}{108} (-19 z-37)\bigg\}
\nn\\
&~~~~~~
+C_F^2 \bigg\{\frac{1}{4} \left(-15 \zeta_3+\frac{511}{16}
-\frac{67 \pi ^2}{12}+\frac{17 \pi ^4}{45}\right) \delta(1-z)
\nn\\
&~~~~~~
+\frac{1+z^2}{1-z}
 \bigg(\frac{\text{Li}_3(1-z)}{2}-\frac{5 \text{Li}_3(z)}{2}+\frac{1}{2} \text{Li}_2(z) \log (1-z)
+\frac{3 \text{Li}_2(z) \log (z)}{2}
\nn\\
&~~~~~~
+\frac{3}{4} \log (z) \log ^2(1-z)+\frac{1}{4} \log ^2(z) \log (1-z)-\frac{1}{12} \pi ^2 \log (1-z)+\frac{5 \zeta_3}{2}\bigg)
\nn\\
&~~~~~~
+(1-z) \left(-\text{Li}_2(z)-\frac{3}{2} \log (1-z) \log (z)+\frac{2 \pi ^2}{3}-\frac{29}{4}\right)
+\frac{1}{24} \left(1+z\right) \log ^3(z)
\nn\\
&~~~~~~
+\frac{1}{1-z}\bigg(
\frac{1}{8} \left(-2 z^2+2 z+3\right) \log ^2(z)+\frac{1}{4} \left(17 z^2-13 z+4\right) \log (z)\bigg)
-\frac{z}{4}\log (1-z)
\bigg\}
\nn\\
&~~~~~~
+C_F \bigg\{\frac{1}{z}(1-z) \left(2 z^2-z+2\right) \left(\frac{\text{Li}_2(z)}{6}+\frac{1}{6} \log (1-z) \log (z)-\frac{\pi ^2}{36}\right)
\nn\\
&~~~~~~
+\frac{1}{216 z}(1-z) \left(136 z^2-143 z+172\right)-\frac{1}{48} \left(8 z^2+3 z+3\right) \log ^2(z)
\nn\\
&~~~~~~
+\frac{1}{36} \left(32 z^2-30 z+21\right) \log (z)+\frac{1}{24} (1+z) \log ^3(z)
\bigg\}\;,
\end{align}
\begin{align}
\label{h2qqbp}
{\cal H}^{DY(2)}_{q\bar q\ito q \bar{q}'}(z)&=
C_F \bigg\{
\frac{1}{12 z}(1-z) \left(2 z^2-z+2\right) \bigg(
\text{Li}_2(z)
+\log (1-z) \log (z)
-\frac{\pi^2}{6}
\bigg)
\nn\\
&~~~~~~
+\frac{1}{432 z}(1-z) \left(136 z^2-143 z+172\right)
+\frac{1}{48} (1+z) \log^3(z)
\nn\\
&~~~~~~
-\frac{1}{96} \left(8 z^2+3 z+3\right) \log^2(z)
+\frac{1}{72} \left(32 z^2-30 z+21\right) \log (z)
\bigg\}\;,
\end{align}
\newpage
\begin{align}
\label{h2qq}
{\cal H}^{DY(2)}_{q\bar q\ito q q}(z)&=
C_F\left(C_F-\frac{1}{2}C_A\right)
 \bigg\{ \frac{1+z^2}{1+z}
\bigg(\frac{3 \text{Li}_3(-z)}{2}+\text{Li}_3(z)
+\text{Li}_3\left(\frac{1}{1+z}\right)-\frac{\text{Li}_2(-z) \log (z)}{2}
\nn\\
&~~~~~~
-\frac{\text{Li}_2(z) \log (z)}{2}-\frac{1}{24} \log ^3(z)
-\frac{1}{6} \log ^3(1+z)+\frac{1}{4} \log (1+z) \log ^2(z)
\nn\\
&~~~~~~
+\frac{\pi^2}{12}  \log (1+z)-\frac{3 \zeta_3}{4}\bigg)
+\left(1-z\right) \left(\frac{\text{Li}_2(z)}{2}+\frac{1}{2} \log (1-z) \log (z)+\frac{15}{8}\right)
\nn\\
&~~~~~~
-\frac{1}{2}(1+z) \big(\text{Li}_2(-z)+ \log (z) \log (1+z)\big)
+ \frac{\pi^2}{24}  (z-3)+\frac{1}{8} (11 z+3) \log (z)\bigg\}
\nn\\
&~~~~~~
+C_F \bigg\{\frac{1}{12 z}(1-z) \left(2 z^2-z+2\right) \left(\text{Li}_2(z)+\log (1-z) \log (z)
-\frac{\pi ^2}{6}\right)
\nn\\
&~~~~~~
+\frac{1}{432 z}(1-z) \left(136 z^2-143 z+172\right)-\frac{1}{96} \left(8 z^2+3 z+3\right) \log ^2(z)
\nn\\
&~~~~~~
+\frac{1}{72} \left(32 z^2-30 z+21\right) \log (z)+\frac{1}{48} (1+z) \log ^3(z)\bigg\}
\;,
\end{align}
%
\begin{align}
\label{h2qqp}
{\cal H}^{DY(2)}_{q\bar q\ito q q'}(z)={\cal H}^{DY(2)}_{q\bar q\ito q \bar{q}'}(z)\;,
\end{align}
\begin{align}
\label{h2qg}
{\cal H}^{DY(2)}_{q \bar q\ito qg}(z)
&=
C_A \bigg\{-\frac{1}{12 z}(1-z) \left(11 z^2-z+2\right) \text{Li}_2(1-z)
\nn\\
&~~~~~~
+\left(2 z^2-2 z+1\right) \bigg(\frac{\text{Li}_3(1-z)}{8}
-\frac{1}{8} \text{Li}_2(1-z) \log (1-z)+\frac{1}{48} \log^3(1-z)\bigg)
\nn\\
&~~~~~~
+\left(2 z^2+2 z+1\right) \bigg(\frac{3 \text{Li}_3(-z)}{8}
+\frac{\text{Li}_3\left(\frac{1}{1+z}\right)}{4}-\frac{\text{Li}_2(-z) \log(z)}{8}
-\frac{1}{24} \log^3(1+z)
\nn\\
&~~~~~~
+\frac{1}{16} \log^2(z) \log (1+z)
+\frac{1}{48} \pi^2 \log (1+z)\bigg)
+\frac{1}{4} z (1+z) \text{Li}_2(-z)+z \text{Li}_3(z)
\nn\\
&~~~~~~
-\frac{1}{2} z \text{Li}_2(1-z) \log(z)-z \text{Li}_2(z) \log(z)
-\frac{3}{8} \left(2 z^2+1\right) \zeta_3-\frac{149 z^2}{216}
\nn\\
&~~~~~~
-\frac{1}{96} \left(44 z^2-12 z+3\right) \log^2(z)
+\frac{1}{72} \left(68 z^2+6 \pi^2 z-30 z+21\right) \log(z)
+\frac{\pi^2 z}{24}+\frac{43 z}{48}
\nn\\
&~~~~~~
+\frac{43}{108 z}
+\frac{1}{48} (2 z+1) \log^3(z)
-\frac{1}{2} z \log (1-z) \log^2(z)
-\frac{1}{8} (1-z) z \log^2(1-z)
\nn\\
&~~~~~~
+\frac{1}{4} z (1+z) \log (1+z) \log(z)
+\frac{1}{16} (3-4 z) z \log (1-z)-\frac{35}{48}\bigg\}
\nn\\
&~~~~~~
+C_F \bigg\{\left(2 z^2-2 z+1\right) 
\bigg(\zeta_3-\frac{\text{Li}_3(1-z)}{8}
-\frac{\text{Li}_3(z)}{8}+\frac{1}{8} \text{Li}_2(1-z) \log (1-z)
\nn\\
&~~~~~~
+\frac{\text{Li}_2(z) \log(z)}{8}-\frac{1}{48} \log^3(1-z)
+\frac{1}{16} \log(z) \log^2(1-z)+\frac{1}{16} \log^2(z) \log (1-z)
\bigg)
\nn\\
&~~~~~~
-\frac{3 z^2}{8}-\frac{1}{96} \left(4 z^2-2 z+1\right) \log^3(z)
+\frac{1}{64} \left(-8 z^2+12 z+1\right) \log^2(z)
\nn\\
&~~~~~~
+\frac{1}{32} \left(-8 z^2+23 z+8\right) \log(z)+\frac{5}{24} \pi^2 (1-z) z
+\frac{11 z}{32}+\frac{1}{8} (1-z) z \log^2(1-z)
\nn\\
&~~~~~~
-\frac{1}{4} (1-z) z \log (1-z) \log(z)
-\frac{1}{16} (3-4 z) z \log (1-z)-\frac{9}{32}\bigg\}\;,
\end{align}
\begin{align}
{\label{h2gg}
\cal H}^{DY(2)}_{q\bar q\ito gg}(z)&=
-\,\f{z}{2}\,\left(\,1-z+\frac{1}{2}\,(1+z)\,\log (z)\, \right)\;,
\end{align}
where 
$C_F=(N_c^2-1)/(2N_c), \,C_A=N_c$ ($N_c$ is the number of colours in $SU(N_c)$
QCD), $n_F$ is the number of quark flavours and
${\rm Li}_k(z)$ $(k=2,3)$ are the usual
polylogarithm functions,
\beq
{\rm Li}_2(z)= - \int_0^z \f{dt}{t} \;\ln(1-t) \;\;,
\quad \quad {\rm Li}_3(z)=  \int_0^1 \f{dt}{t} \;\ln(t)\;\ln(1-zt) \;\;.
\eeq

We comment on the vector boson results in Eqs.~(\ref{h2qqb})--(\ref{h2gg}) 
and on the ensuing determination
of the second-order 
coefficients 
$C_{q\,q}^{(2)}$, $C_{q\,q'}^{(2)}$,
$C_{q\,{\bar q}}^{(2)}$, $C_{q\,{\bar q}'}^{(2)}$,
 $C_{q\,g}^{(2)}$
and $H_q^{DY(2)}$
in Eqs.~(\ref{cpert})
and  (\ref{hpert}).

The parton matrix ${\cal H}^{DY(2)}_{q\bar q\ito ab}$ is completely specified
by the six entries\footnote{The other non-vanishing entries are obtained by the
symmetry relation 
${\cal H}^{DY}_{q\bar q\ito ab}= 
 {\cal H}^{DY}_{q\bar q\ito {\bar b}{\bar a}}$. Several entries of the
 second-order matrix ${\cal H}^{DY (2)}_{q\bar q\ito ab}$ are vanishing because
 of Eq.~(\ref{cone0}).} 
in Eqs.~(\ref{h2qqb})--(\ref{h2gg}): the quark--quark
functions ${\cal H}^{DY(2)}_{q \bar q\ito q \bar q}$,  
${\cal H}^{DY(2)}_{q \bar q\ito q \bar{q}'}$, 
${\cal H}^{DY(2)}_{q \bar q\ito q q}$, ${\cal H}^{DY(2)}_{q \bar q\ito q q'}$, 
the quark--gluon
function ${\cal H}^{DY(2)}_{q \bar q\ito qg}$ and  the gluon--gluon
function ${\cal H}^{DY(2)}_{q \bar q\ito gg}$.

Using Eq.~(\ref{H2}), in the gluon--gluon channel we have
\begin{equation}
\label{gg2}
{\cal H}^{DY(2)}_{q\bar q\ito gg}(z)=
\left(C^{(1)}_{q\,g}\otimes C^{(1)}_{q\,g}\right)(z) \;\; .
\end{equation}
We see that the second-order coefficient function 
${\cal H}^{DY(2)}_{q\bar q\ito gg}(z)$ is fully determined by the $q_T$
resummation coefficients at ${\cal O}(\as)$. Using the value of 
$C^{(1)}_{q\,g}$ in Eq.~(\ref{coneqg}),
the expression on the right-hand side of Eq.~(\ref{gg2}) is in complete 
agreement with the result in Eq.~(\ref{h2gg}).
Therefore, our explicit computation of the NNLO partonic function
${\hat R}^{V (2)}_{q\bar q\ito gg}$ represents a consistency check of the 
resummation formula (\ref{qtycrossgg}).

Considering the quark--gluon channel, Eq.~(\ref{H2}) can be recast in the
following form:
\begin{equation}
\label{qg2}
C^{(2)}_{q\,g}(z) + \f{1}{2} \,H^{DY(1)}_q \,C^{(1)}_{q\,g}(z) =
{\cal H}^{DY(2)}_{q\bar q\ito qg}(z) 
- \f{1}{2} \left( {\cal H}^{DY(1)}_{q\bar q\ito q \bar q} \otimes C^{(1)}_{q\,g}
\right)(z)
\;\; ,
\end{equation}
where we have used
${\cal H}^{DY(1)}_{q\bar q\ito q\bar q}(z) = H^{DY(1)}_q \,\delta(1-z) 
+ 2 \,C^{(1)}_{q\,q}(z)$ (see Eq.~(\ref{H1})).
The relation (\ref{qg2}) can be used to determine $C^{(2)}_{q\,g}(z)$ from the
knowledge of ${\cal H}^{DY(2)}_{q\bar q\ito qg}$ and of the 
$q_T$ resummation coefficients at ${\cal O}(\as)$.
Inserting the first-order results of Eqs.~(\ref{coneqg})--(\ref{coneqq})
in Eq.~(\ref{qg2}), we explicitly have
\begin{equation}
\label{qg22}
C^{(2)}_{q\,g}(z) + \f{1}{4} \,H^{DY(1)}_q \,z\,(1-z) =
{\cal H}^{DY(2)}_{q\bar q\ito qg}(z) - \f{C_F}{4} \left[
z \log (z)+\frac{1}{2}\,(1-z^2)+\left( \frac{\pi^2}{2}-4\right)z\,(1-z)
\right] \;,
\end{equation}
where ${\cal H}^{DY(2)}_{q\bar q\ito qg}$ is given in Eq.~(\ref{h2qg}).
Note that the right-hand side of Eq.~(\ref{qg2}) (or Eq.~(\ref{qg22}))
is resummation-scheme independent. Analogously to Eq.~(\ref{coneqq}),
the dependence of $C^{(2)}_{q\,g}$ on the resummation scheme is thus parametrized
by the first-order coefficient $H^{DY(1)}_q$ on the left-hand side of 
Eq.~(\ref{qg22}).

The process-independent coefficient functions  $C^{(2)}_{q\,q}(z)$, $C^{(2)}_{q\,q'}(z)$ 
$C^{(2)}_{q\,\bar q}(z)$ and $C^{(2)}_{q\,\bar{q}'}(z)$ are obtained
analogously to $C^{(2)}_{q\,g}(z)$. Considering the flavour diagonal quark--quark channel,
Eq.~(\ref{H2}) gives
\begin{align}
\label{qq2}
2 \,C^{(2)}_{q\,q}(z) &+ \delta(1-z) \left[ H^{DY(2)}_q - \f{3}{4} 
\left(H^{DY(1)}_q\right)^2 \right] +
\f{1}{2} \,H^{DY(1)}_q \,{\cal H}^{DY(1)}_{q\bar q\ito q\bar q}(z) \nn \\
&={\cal H}^{DY(2)}_{q\bar q\ito q\bar q}(z) 
- \f{1}{4} \left( {\cal H}^{DY(1)}_{q\bar q\ito q\bar q} \otimes 
{\cal H}^{DY(1)}_{q\bar q\ito q\bar q}\right)(z)\;\; ,
\end{align}
where the right-hand side of Eq.~(\ref{qq2}) is expressed in terms of resummation-scheme 
independent functions. Inserting Eqs.~(\ref{coneqg})--(\ref{coneqq}) in 
Eq.~(\ref{qq2}), we explicitly obtain
\begin{align}
\label{qq22}
2 \,C^{(2)}_{q\,q}(z) &+ \delta(1-z) \left[ H^{DY(2)}_q - \f{3}{4} 
\left(H^{DY(1)}_q\right)^2 + \f{C_F}{4}\,(\pi^2-8) \,H^{DY(1)}_q \right] +
\f{1}{2}C_F\,H^{DY(1)}_q (1-z)
 \nn \\
&={\cal H}^{DY(2)}_{q\bar q\ito q\bar q}(z) 
- \f{C_F^2}{4}\left[\, \delta(1-z)  \f{(\pi^2-8)^2}{4}
+ \left(\pi^2-10\right)(1-z) - (1+z) \ln z\, \right] \;\; ,
\end{align}
where ${\cal H}^{DY(2)}_{q\bar q\ito q\bar q}$ is given in Eq.~(\ref{h2qqb}).
We observe that $C^{(2)}_{q\,q}(z)$ includes a resummation-scheme dependent part
that depends on 
$H^{DY(1)}_q$ and $H^{DY(2)}_q$. We also recall \cite{Catani:2000vq} that the 
resummation-scheme invariance relates $C^{(2)}_{q\,q}$, $H^{DY(2)}_q$ and the
third-order coefficient $B^{(3)}_q$ of the quark form factor.

Considering the flavour off-diagonal quark--quark channel in
Eq.~(\ref{H2}), we obtain
\begin{align}
\label{qqp2}
C^{(2)}_{q\,\bar q}(z)=&{\cal H}^{DY(2)}_{q\bar q\ito q q}(z)\;\; ,\quad
C^{(2)}_{q\,q'}(z)={\cal H}^{DY(2)}_{q\bar q\ito q \bar{q}'}(z)\;\; ,\quad
C^{(2)}_{q\,\bar{q}'}(z)={\cal H}^{DY(2)}_{q\bar q\ito q q'}(z)\;\; ,
\end{align}
where ${\cal H}^{DY(2)}_{q\bar q\ito q \bar{q}'}$, ${\cal H}^{DY(2)}_{q\bar q\ito q q}$, 
and ${\cal H}^{DY(2)}_{q\bar q\ito q q'}$ are given in Eqs.~(\ref{h2qqbp})--(\ref{h2qqp}).
The off-diagonal second-order coefficients $C^{(2)}_{q\,\bar q}(z)$, 
$C^{(2)}_{q\,q'}(z)$ 
and $C^{(2)}_{q\,\bar{q}'}(z)$ are resummation-scheme independent.
From Eq.~(\ref{h2qqp}) we observe that we have 
$C^{(2)}_{q\,q'}(z)=C^{(2)}_{q\,\bar{q}'}(z)$. 
The equality between $C_{q\,q'}(z)$ and $C_{q\,\bar{q}'}(z)$ 
is expected to be violated at higher perturbative orders (i.e., we expect
$C^{(3)}_{q\,q'}(z) \neq C^{(3)}_{q\,\bar{q}'}(z)$).

In this paper we have considered
QCD radiative corrections to vector boson production 
in hadron--hadron collisions. 
We have presented the analytic result of the NNLO calculation of the vector
boson cross 
section at small values of $q_T$ 
(see Eqs.~(\ref{inte}) and (\ref{eqr2s})).
The NNLO result is compared 
(see Eq.~(\ref{eqr2}))
with the predictions 
of transverse-momentum resummation.
The comparison gives a second-order crosscheck of the all-order
resummation formula (\ref{qtycrossgg}),
and it allows us to determine the previously unknown resummation coefficients 
at ${\cal O}(\as^2)$.
These are the coefficient functions ${\cal H}^{DY(2)}_{q\bar q\ito ab}(z)$
(see Eqs.~(\ref{h2qqb})--(\ref{h2gg}))
and the related coefficients $C^{(2)}_{q\,g}$, $C^{(2)}_{q\,q}$,
$C^{(2)}_{q\,q'}(z)$,
$C^{(2)}_{q\,\bar q}(z)$ and $C^{(2)}_{q\,\bar{q}'}(z)$
(see Eqs.~(\ref{qg22}), (\ref{qq22}) and (\ref{qqp2})), 
which control the dependence on the rapidity
of the vector boson.
The knowledge of these second-order coefficients is relevant for
phenomenological applications of both resummed and fixed-order QCD computations.
These coefficients have been already implemented in resummed calculations of the inclusive
$q_T$ distribution at full NNLL accuracy \cite{Bozzi:2010xn}.
Using the method of Ref.~\cite{Catani:2007vq},
the same coefficients 
have been used to perform the fully-exclusive NNLO
perturbative calculations of Refs.~\cite{Catani:2009sm} and 
\cite{Ferrera:2011bk}.

\noindent {\bf Acknowledgements.}
This work was supported in part by UBACYT, CONICET, ANPCyT, INFN and
the Research Executive Agency (REA) 
of the European Union under the Grant Agreement number PITN-GA-2010-264564 
({\it LHCPhenoNet},  Initial Training Network).

\end{document}